# Quantum engineering of transistors based on 2D materials heterostructures

Giuseppe Iannaccone[1,*], Francesco Bonaccorso[2,*], Luigi Colombo[3], Gianluca Fiori[1]

[1] *Dipartimento di Ingegneria dell'Informazione, Università di Pisa, 56122 Pisa, Italy;*

[2]*Istituto Italiano di tecnologia, Graphene Labs, 16163 Genova, Italy;*

[3]*University of Texas at Dallas, Richardson, Texas 75080, USA.*

## Abstract

Quantum engineering entails atom by atom design and fabrication of electronic devices. This innovative technology that unifies materials science and device engineering has been fostered by the recent progress in the fabrication of vertical and lateral heterostructures of two-dimensional materials and by the assessment of the technology potential via computational nanotechnology. But how close are we to the possibility of practical realisation of the next generation atomically thin transistors? In this perspective we analyse the outlook and the challenges of quantum-engineered transistors using heterostructures of two-dimensional materials against the benchmark of silicon technology and its foreseeable evolution in terms of potential performance and manufacturability. Transistors based on lateral heterostructures emerge as the most promising option from a performance point of view, even if heterostructure formation and control are in the initial technology development stage.

## Introduction

Recently, the so-called "materials-on-demand paradigm" has been proposed [1] thanks to the possibility of forming a "three-dimensional (3D) material" with tailored characteristics by combining layers of two-dimensional (2D) materials, such as graphene [2], and other single/few-atomic-layer semiconductors, insulators, and metals [3][4]. In many ways, this paradigm is a modern and challenging evolution of what in the 1980s was called "band-gap engineering" [5] or "band-structure engineering" [6], *i.e.*, the artificial modification of band edge profiles using heterostructures made possible by epitaxial growth of III-V, II-VI, and IV-IV [7] material systems. Pioneering the use of heterostructures of nearly lattice-matched semiconductor layers of few nanometres, researchers at US and Japanese research laboratories proposed and demonstrated new device concepts, such as hot electron transistors [8] or resonant tunnelling transistors [9][10], as well as optimized devices, such as graded-base bipolar transistors [11]. This capability transformed optoelectronics and electronics, as testified by the realization of the fastest transistor commonly used in telecommunications, *i.e.*, the heterojunction bipolar transistor.

The Semiconductor Industry has made intense use of new materials and device engineering techniques in order to sustain Moore's law. This has been especially true in the last fifteen years, when strained silicon (using SiGe in the source and drain) was introduced at the so-called "90 nm" node, high-k gate dielectrics and metal gates at the 45 nm node, and the tri-gate geometries at the 22 nm node [12]. As devices are scaled down to nanometre feature sizes, the boundary between materials and devices as separate physical abstractions become vanishingly small.

Lateral and vertical heterostructures of 2D materials can represent an enabling device engineering technology beyond what can be accomplished by either Si and SiGe, or III-V materials systems (cubic systems). Indeed, the ability to fabricate lateral 2D heterostructures with graphene and hexagonal boron nitride (*h*-BN) [13][14][15][16], or with different phases (metallic and semiconducting) of transition metal mono- and di-chalcogenides (TMDs) [17][18] provides additional degrees of freedom for device engineering at the atomistic scale, with vertical and/or planar heterostructures. *These developments have the potential to revolutionize electronics and optoelectronics via quantum engineering of electron devices.* Several recent reports suggested [19][20] that twisted layers of 2D materials can modify the transport properties of the stacked layers significantly. This method of forming heterostructures using 2D materials can significantly modify the band-structure and hence transport properties [19][20][21].

There is a difference in scope between quantum engineering and band structure engineering: while the latter mainly enables a modulation of the band edge profiles and of bandgaps by controlling alloy composition during materials growth, quantum engineering involves heterostructures with layers that often have no atomic species in common (*e.g.*, $MoS_2$-$WSe_2$), or have incommensurate lattice and completely different band structure. In addition, combinations of lateral and vertical heterostructures can be devised, largely expanding the range of possibilities in the direction of atom-by-atom transistor design.

Pioneering research has already demonstrated first proof-of-principle transistors with vertical graphene-based [22][23][24][25][26] or TMD heterostructures [27][28], transistors with lateral heterostructures [29][30][31][32] and non-volatile memory cells [33], as well as optoelectronic devices such as photodiodes based on 2D vertical heterostructures, [35][36] and high quantum efficiency photovoltaic cells based on TMDs/graphene stacks [37]. The number of 2D materials already used to form the so-called "van der Waals heterostructures" is of the order of a few tens [3]. This number will surely increase as the number of 2D materials and more conventional heterostructures or metal/semiconductor contacts are identified and fundamental properties studied.

In this perspective article we discuss the challenges, the opportunities, and the potential of quantum engineered transistors by exploiting the fundamental properties of 2D materials and their heterostructures. We focus on field-effect transistors (FETs), because they can be used with current digital logic architectures and already exhibit promising operation at room temperature. Devices based on other operation mechanisms [38][39] are beyond the scope of the present paper.

We consider the fundamental issues related to device/material fabrication from an industrial point of view, and the attainable device performance, using as a benchmark the expected evolution of complementary metal-oxide semiconductor (CMOS) technology, outlined by Semiconductor Industry roadmaps [40][41]. We also critically discuss transistor structures that - even under extremely optimistic fabrication conditions - cannot be competitive with the foreseeable evolution of CMOS technology, *i.e.*, the continued scaling down of FinFETs below the present-day 14-nm technology node and further scaled gate-all-around FETs.

Vertical stacking or lateral growth of different 2D materials is a new and very complex challenge. While stacking of 2D materials for vertical heterostructures presents the issue of registration or orientation control of one layer to the other, lateral growth requires "lateral lattice matching", which is still at the embryonic stage of development. These challenges also present opportunities that both material scientists and device engineers must address to optimize device design and performance.

## Fabrication of vertical and lateral heterostructures

The direct growth of III-V, II-VI, and IV-IV heterostructures has given rise to many devices today in volume manufacturing [45]. Heterostructures of these materials have been grown predominantly by molecular beam epitaxy (MBE) [46], and by metal-organic chemical vapour deposition and, given that the individual layers have a thickness of at least several nanometres, their growth at reduced temperatures, to minimize inter-diffusion, is now pervasive.

The fabrication of 2D materials-based heterostructures presents a new set of opportunities and challenges. A key aspect of 2D materials is the fact that under ideal conditions they are expected to have low interfacial defects unlike 3D materials, and, because of the interlayer weak van der Waals bonding, they are also expected to have lower inter-diffusion compared to their 3D counterparts. Moreover, 2D materials-based vertical heterostructures are less sensitive to lattice mismatch thus enabling high-quality abrupt interfaces with low trap densities. Additionally, perhaps a minor point, less material will be needed for device fabrication with respect to 3D cubic materials.

Significant progress has been made in the growth of graphene on metals and on SiC [47]. Progress has also been reported on the growth of *h*-BN [48][49] and TMD [50][51][53][52] materials, but growth of large area monolayer or controlled few-layer single crystals is still elusive and will require significant time and resources. As with any new material, there are many challenges in the deposition of 2D materials to form heterostructures, depending on the specific 2D material. For example, in the case of graphene and *h*-BN, it is important to use a catalyst to grow the film [48][49][50][54] and the challenge is to transfer the film on the desired substrate or to grow it *in-situ*. In the case of TMD materials family, the challenge is in the nucleation and growth of single crystals on dissimilar surfaces (dielectric oxides, TMDs, or *h*-BN).

Lateral growth of 2D materials also presents a number of opportunities. In this case, the structures must be fabricated by creating materials sequentially by selective lateral growth processes. The development of this technology may also enable bottom-up techniques [53]. Two-dimensional growth has already been clearly demonstrated for graphene on Cu [54]. Deterministic nucleation and growth of graphene on Cu has also been proven to yield hexagonal graphene single crystals [55]. In principle, this process can be extended to the growth of lateral structures that follow predetermined layouts. The process of course will require careful edge functionalization depending upon lateral heterostructure composition profile. The growth anisotropy is a critical advantage of 2D materials in comparison to 3D materials, which has already been demonstrated in both graphene [54], [56] and more recently in TMDs [57].

The most important challenge, for devices based on both lateral and vertical heterostructures, is the preparation of the basic materials followed by their integration in the desired device structure. While vertical heterostructures, in principle, can be formed by direct growth (Fig. 1a) or by a transfer process for planar geometries (Fig. 1b), lateral heterostructures must be grown in place by some chemical means. Currently, three methods are used or under evaluation for the fabrication of 2D materials-based heterostructures [58]: (*i*) layer-by-layer stacking via mechanical transfer of CVD grown films and exfoliated natural or synthetic bulk grown 2D materials; (*ii*) direct growth by CVD, MBE, or atomic layer epitaxy (ALE); and (*iii*) layer-by-layer deposition of solution-processed 2D crystals. However, at this time all of the aforementioned approaches have limitations.

Layer-by-layer stacking or deterministic placement of 2D materials via mechanical transfer is the only technique that has been used to create heterostructures. The method has relied principally on the mechanical exfoliation of bulk layered materials into atomically thin sheets [59], but more recently the direct transfer of CVD-grown films (Fig. 1b) has also been used. The method has been extensively used for graphene, bi-layer graphene, *h*-BN, and TMDs to fabricate various stacked devices. While transfer techniques of synthetic films or exfoliated layers from natural or synthetic bulk crystals have enabled the demonstration of many proof-of-principle stacked devices, allowing fundamental understanding of physical phenomena in these structures, manufacturing processes of heterostructures on a large scale have not been developed yet. It should be noted that mechanical exfoliation from either natural or synthetic bulk crystals at this time is not believed to be a manufacturable process, given the small size of the available crystals. So far, CVD or MBE techniques have been favoured for the growth of thin films of the various 2D materials [47]; these growth techniques will certainly be optimized for single layer films of graphene, *h*-BN and TMDs, as has been done for traditional materials used by the semiconductor industry.

The selection of the growth technique will depend mainly on device design and integration scheme and secondarily on techniques that yield the highest quality material. The availability of high-quality synthetic 2D films will enable the development of equipment for transfer with and without rotational alignment of 2D materials for the fabrication of any type of planar on-demand vertical heterostructure. As new equipment is developed for transfer and placement of the 2D films, many opportunities will emerge for the integration of these material structures in semiconductor manufacturing flows. The ideal or preferred case would be to use MBE, CVD and ALE processes to create/grow in-situ new heterostructures. However, it is envisioned that the growth of on-demand aligned heterostructures (see Fig.1a) using these techniques could be extremely difficult, particularly in the case of twisted layers (Fig.1c).

In the case of aligned heterostructures, nucleation and growth processes will have to be developed. The growth of these heterostructures on a large scale could be facilitated by the development of selective ALE growth, thus mitigating the need for large area single crystals. However, the case of twisted heterostructures, not discussed here given the immaturity of this technology, will require transfer of individual films with precise rotational alignment on planar surfaces. If *in-situ* growth of layered structures becomes elusive, dry transfer protocols must be developed to create vertical heterostructures.

The manufacturing techniques will have to guarantee the transfer of films with clean surfaces to enable the ultimately needed low-defect-density 2D-2D interfaces. It is important to note that these transfer techniques will be limited to planar vertically aligned and twisted structures [3].

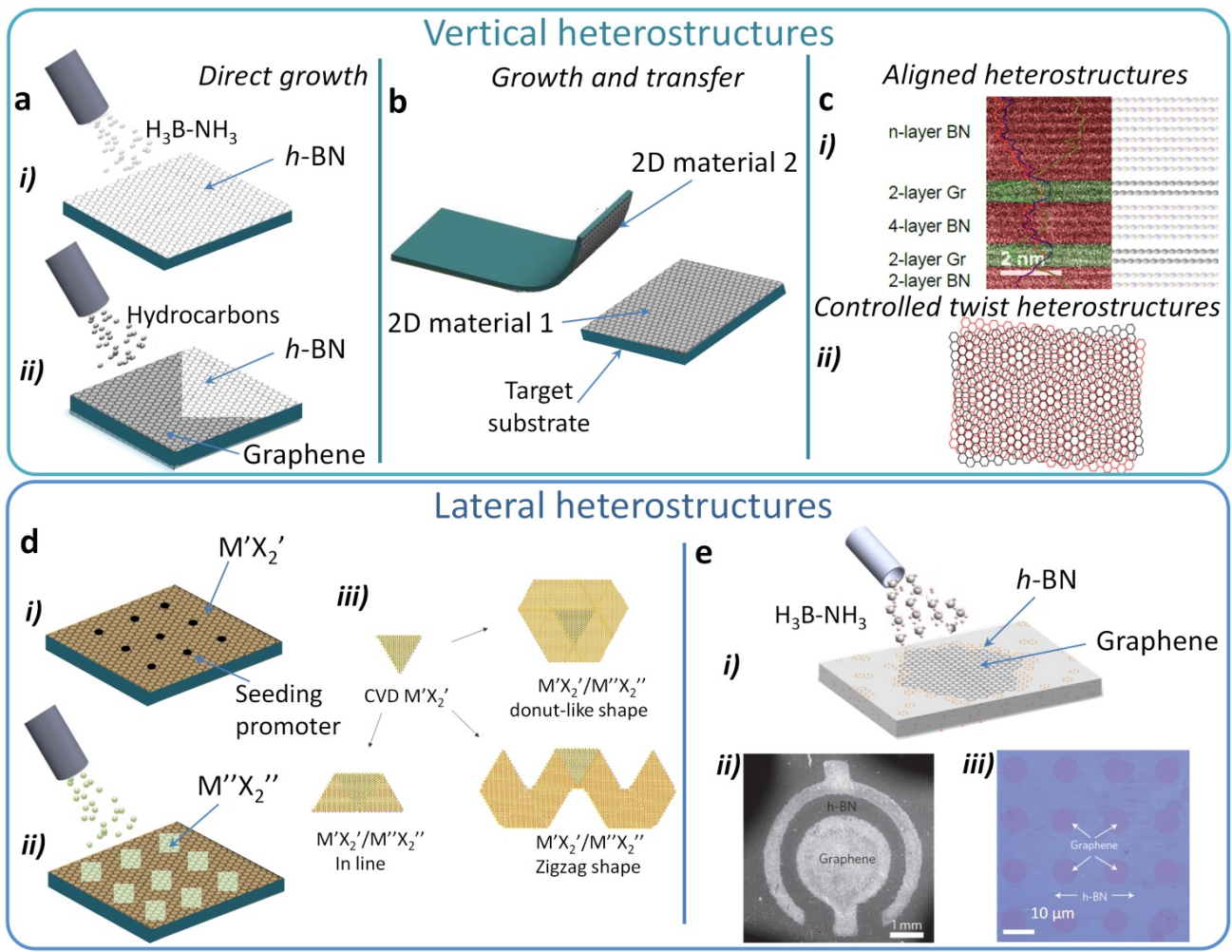

*Figure 1:* a) Direct growth of 2D-materials-based vertical heterostructures by CVD with (i) h-BN by exploiting ammonia borane as precursor and ii) graphene onto the as-grown h-BN layers. b) Growth by either CVD or MBE of individual 2D materials and subsequent dry transfer using pick and place techniques enabling, in principle, any combination of different 2D materials. c) The relative orientation of the different layers of 2D materials is key and mandatory to be controlled to design i) vertically aligned and ii) controlled twist heterostructures. d) Lateral heterostructures can be realized i) by seeding an already grown 2D material template, ii) grow a second 2D material by using the appropriate precursors; and iii) by a proper placement of seeds through either a pattern and etch process or a mask, which can allow the realization of different lateral heterostructures such as linear, zigzag and donut-like shape. e) i) Direct growth of h-BN on graphene edges; ii) Scanning electron microscopy image showing a concentric h-BN/graphene heterostructure; iii) Optical image of a graphene/h-BN array of circles, with graphene circles embedded in an h-BN matrix. Panel i) in c) Courtesy of Profs. M. Kim and E. Tutuc, panel ii) and iii) in e) adapted from Ref. [15].

Recent reports [56][61][62] on dry transfer of graphene using pick and place techniques exploiting *h*-BN as the dielectric has proved to be successful through the demonstration of very high carrier mobility (ranging from 30,000 *to* 80,000 cm$^2$V$^{-1}$s$^{-1}$) in graphene on *h*-BN. These results suggest that high quality films can be achieved; the question is how these processes can be transferred to a manufacturing environment. In the case of *h*-BN, very small high-quality films can be exfoliated from small bulk synthetic crystals but - while progress is being made in growing large-area thin films by CVD processes - the quality is still not as high as the exfoliated films and the thin film crystals are still at the micron level. Therefore, more work has to be done to achieve high quality synthetic *h*-BN thin films. Additionally, if rotational alignment of 2D material heterostructures begins to yield devices with performance that exceeds non-controlled stacking or registered films, the industry will have to develop tools to achieve controlled alignment of 2D materials.

The growth of TMDs face similar challenges as *h*-BN in that wafer-scale single crystals are still not available and stacked-crystals, layer-by-layer wafer-scale single crystal growth has yet to be demonstrated. Significant efforts are being dedicated to the understanding of single crystal growth by chemical vapour deposition through simulations and experiments [51][64]. The growth of lateral heterostructures (see Fig. 1 d,e) as described above will require a different approach. Seeding experiments performed up until now [64][65] will form the basis for future lateral heterostructures. Seeding of both graphene and TMD films has already been demonstrated and it is not difficult to imagine a process where one can grow sequential films in a similar way as isotopic C was used to demonstrate lateral growth of graphene on Cu [54].

For lower performance devices, layer-by-layer deposition from 2D crystal-based inks is another strategy for the scalable production of 2D heterostructures [66][67]. However, although progress has been made in this field [66][67][68][69], even under the best conditions these materials will not meet high-performance device requirements and thus will not be discussed further in this article.

## Vertical and lateral heterostructure field effect transistors

In the pioneering years of a new field in science and technology, creativity flourishes and many original concepts are proposed, sometimes at the expense of good design. After this initial creative phase, a screening phase is needed, since the measure of an electron device is the improvement in performance and in functionality over the incumbent technology.

Indeed, the possibilities opened by vertical and lateral heterostructures have unleashed the creativity of researchers in recent years, leading to the proposals of new transistor concepts and to the experimental demonstration of different heterostructures of 2D materials. Figure 2 shows seven examples of potential transistor structures. At the most basic level, they share a common operating principle: the current between one contact called "source" and the other contact called "drain", separated by an energy barrier – sometimes called "channel" - is controlled by modulating the height or the shape of such barrier via the voltage applied to the "gate" electrode. For most transistor structures, we can have a "lateral" implementation, in which current flows in the plane, through a lateral heterostructure, and a "vertical" implementation, in which current flows in the direction perpendicular to the layers.

In vertical [22][27][24] and lateral [28][30][31][32][70] heterostructure FETs, the channel consists of a material with a larger gap than that used in the source and drain regions. In both cases, the voltage applied to the top gate modulates the energy barrier height and therefore the current. An interesting aspect is that heterostructures enable real device optimization: on the one hand, the large gap region enables the suppression of the current in the OFF state, thus yielding large current modulation; on the other hand, mobility is inversely correlated with the energy gap, *i.e.*, small or zero-gap materials in the source and drain regions provide us with high mobility regions and low-resistance contacts [4].

In vertical [23] and lateral [70] barristors, the Schottky barrier height between a semimetal source with a low density of states (*e.g.*, graphene) and a semiconductor drain is modulated by the top gate voltage, thus modulating the thermionic current.

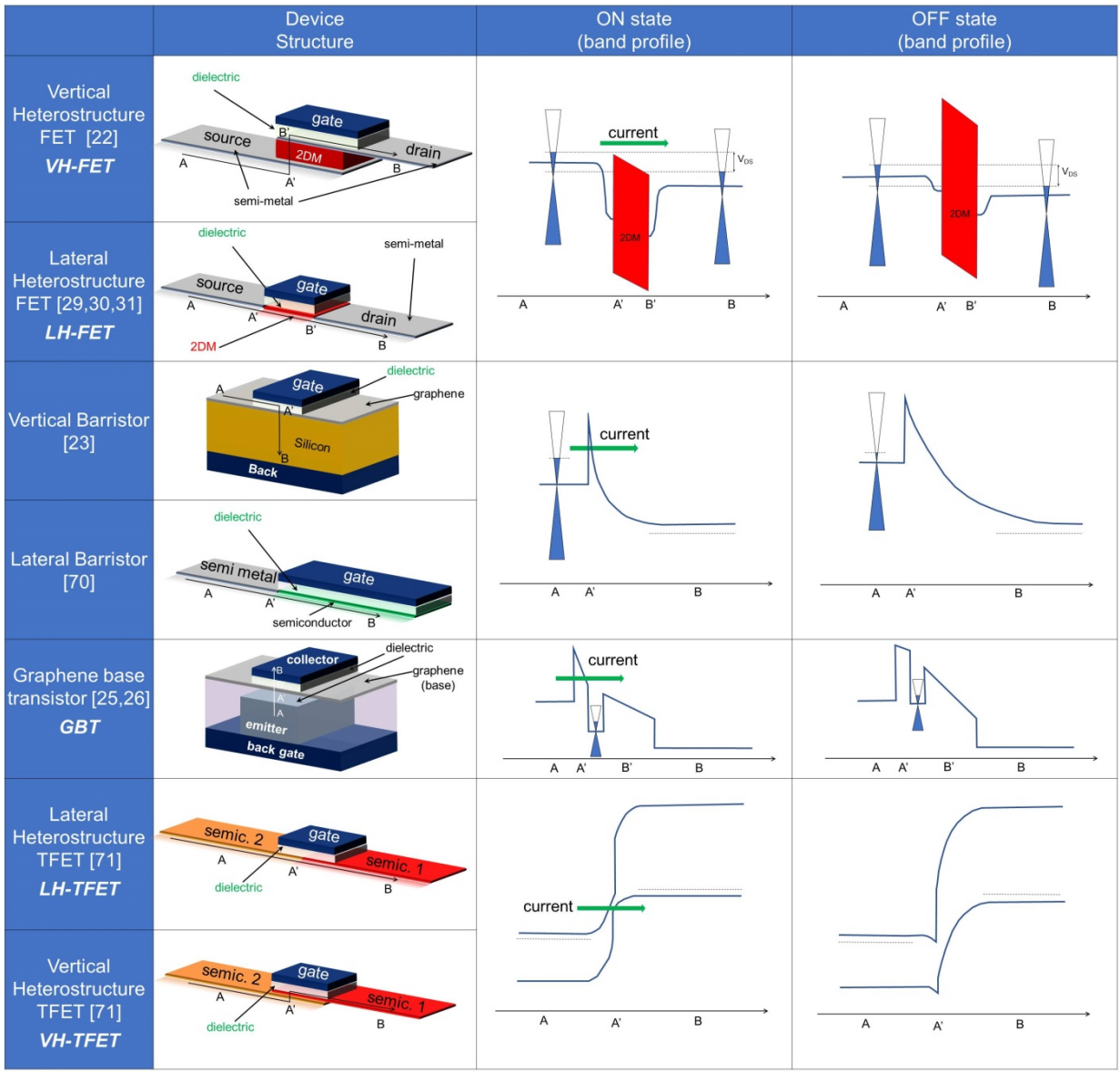

*Figure 2: Seven proposed transistor structures based on 2D heterostructures: first column: illustration of the device structure with main elements highlighted (source, drain, gate, barrier); second column: sketch of the band-edge profile along the transport direction in the ON state; third column: sketch of the band-edge profile along the transport direction in the OFF state.*

In the case of the graphene base transistor [25][26], the barrier between source and drain is represented by a graphene sheet sandwiched between two insulator or semiconducting layers. The voltage applied to the graphene sheet, the "base", modulates the shape and the height of the barrier. This structure is intrinsically vertical, since its main asset is the exploitation of a one-atom-thick base, which – at least in principle – can enable ultrafast traversal times. In practice, radio-frequency operation of graphene base transistors is limited by inter-electrode capacitances, which can dominate over the capacitance associated to charge transport, and by the relatively small tunnelling/thermionic currents.

Another option is represented by vertical and lateral tunnel FETs, in which drain current flows via interband tunnelling through a lateral or a vertical type-II heterostructure [71]. In this case, by varying the voltage applied to the top gate, one can modify the tunnelling barrier profile, hence its transmission coefficient and the current (Fig. 2). Type-II heterostructures are particularly convenient because they can provide a very transparent barrier in the "On" state, due to the staggered band alignment [72].

All the mentioned transistor structures represent a revolution in terms of materials and device structures, but they still must be used in conventional digital logic, *i.e.*, the CMOS logic architectures used in present-day silicon technology. For this reason, we can assess the prospects of each device type using as a benchmark the consensus on the expected evolution of silicon technology illustrated in Fig. 3, reached by the International Technology Roadmap for Semiconductors (ITRS) 2.0, 2015 edition [40], and by its successor, the International Roadmap for Devices and Systems (IRDS) [41].

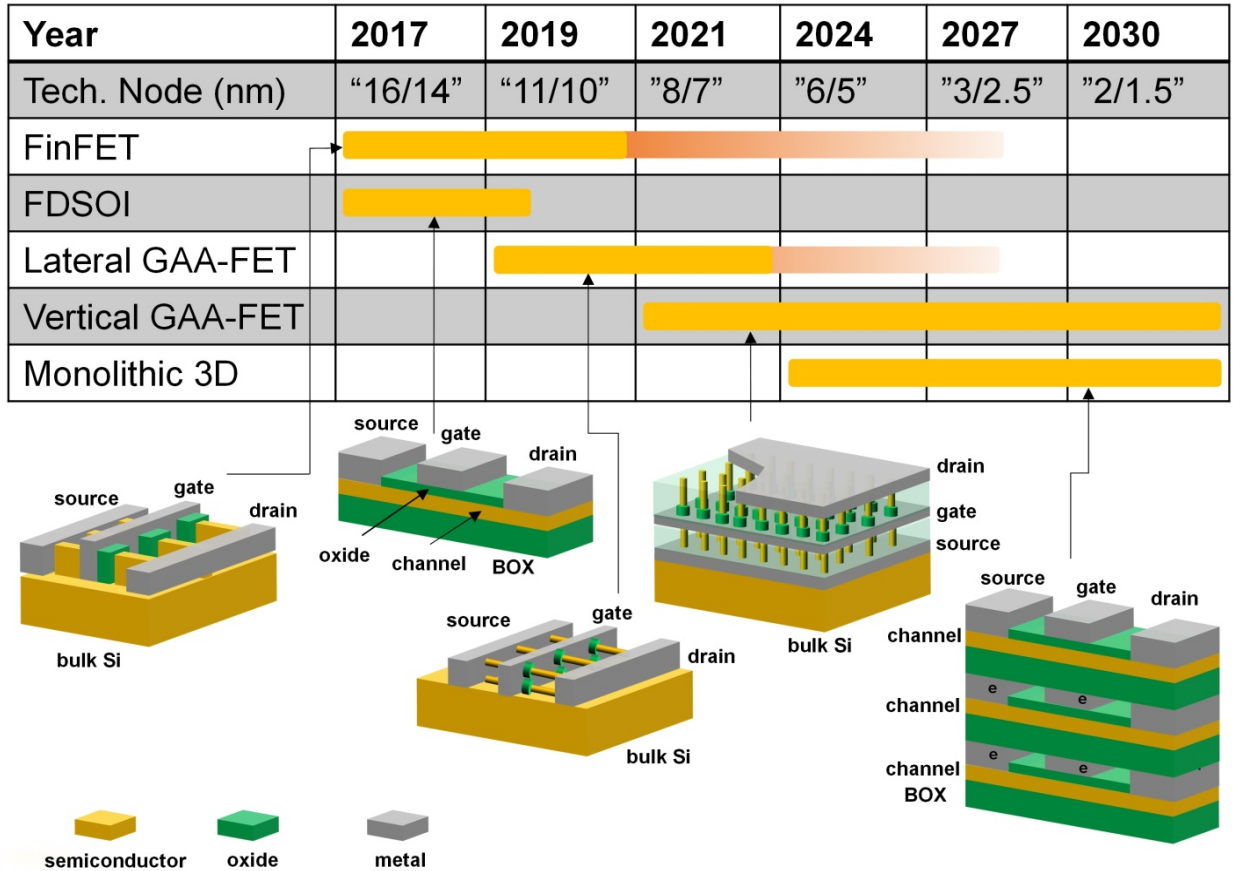

*Figure 3*: Consensus on the transistor device structure to be used in CMOS chips versus year of first shipment, according to the ITRS 2.0, 2015 edition [40], and by its successor, the IRDS [41] (yellow bars). The orange bars refer to the experimental results from [43] on the "7 nm" node FinFETs and to TCAD analysis [44].

According to these consensus documents and to more recent experimental results [43] and technology computer aided design (TCAD) analysis [44], in the short term the so-called Fin FETs, proposed by UC Berkeley in 1998 [42] and manufactured by Intel starting in 2011 for the 22 nm node, and the lateral gate-all around (GAA) FETs can sustain Moore's law up to the so-called "8/7 nm" technology node (expected in 2021) and possibly even beyond based on fabrication considerations [44]. Fully-depleted Silicon-on-Insulator (FDSOI) technology is expected to stop at the "11/10 nm" technology node. After 2021, both the ITRS and the IRDS acknowledge yet unresolved issues, indicating that vertical GAA transistors or monolithic 3D integration, *i.e.*, vertical stacking of planar transistors, will be the likely solutions. Increasing the number of vertically stacked transistors enables the continued increase of the number of transistors per chip, required by Moore's law, even at constant transistor size.

In this context, transistors based on 2D materials can have intrinsic advantages, since the van der Waals interaction between adjacent layers pose less stringent constraints on vertical stacking, and therefore might be suitable to the stacking of many transistor layers with no or limited performance degradation.

Limited interaction between adjacent layers is promising for preserving charge carrier mobility in the case of very thin layers. It is indeed well-known [73][74][75][77] that in the case of silicon, germanium, and InAs charge carrier mobility sharply decreases with decreasing layer thickness. Indeed, with decreasing layer thickness, we have increased quantum confinement of electrons that causes an increase of electron-phonon scattering rates, surface roughness scattering rates, and scattering due to thickness variations and remote phonons, all mechanisms leading to mobility degradation (Fig. 4a). On the contrary, semiconducting TMDs can provide thinner layers with a mobility value in the 20-200 cm$^2$/Vs range [78][79][81][82][84]. Graphene has been shown to achieve mobility at room temperature exceeding 2,000 cm$^2$/Vs on SiO$_2$ substrates [76] and close to 80,000 cm$^2$/Vs on hexagonal BN [61].

The corresponding mean free path (Fig. 4b) in TMD films is in the range from 1 to 4 nm for TMDs and of 20-100 nm for graphene, as extracted from mobility data of Fig. 4a, which is comparable to the channel or barrier region of a 2D heterostructure-based transistor [22][27][24][31][70]. This suggests that transport will be dominated by the quality of the lateral heterojunctions, which are, however, still insufficiently controlled and understood.

Contact resistance ($R_C$) of 2D materials is still an open issue [89] given that present-day transistors have a $R_C < 1\ \Omega \cdot \mu m$ and the best results with graphene are of order $10\ \Omega \cdot \mu m$, [90] and at least one order of magnitude higher for contacts to TMDs [86]. However, the evolution of FETs to ultra-thin body structures reduces the difference between 2D materials and bulk semiconductors (Fig. 4c). In addition, requirements on $R_C$ can likely be relaxed by one order of magnitude for digital applications, and heterostructures allow contact materials optimization. Still, this is one of the most critical areas for the development of electronic devices based on 2D materials.

To assess the transistor performance, the two main figures of merit (FoM) for digital applications are the effective delay time, which is a measure of device speed, and the power-delay product, which is a measure of energy efficiency [58]. The effective delay time is defined as

$$\tau = CV_{DD}/I_{ON} \qquad \text{(Eq. 1)}$$

and the power-delay product as

$$\text{PDP} = V_{DD}I_{ON}\tau = CV_{DD}^2 \qquad \text{(Eq. 2)}$$

where C is the transistor gate capacitance and $V_{DD}$ is the supply voltage.

Here, we should stress that, in a n-type FET, both the so-called OFF current $I_{OFF}$ and ON current $I_{ON}$ are the drain currents when the source is at zero voltage and the drain is at the supply voltage $V_{DD}$. $I_{ON}$ is obtained when the gate is at $V_{DD}$ (ON state), whereas $I_{OFF}$ is obtained when the gate is at zero voltage (OFF state).

The FoM must be referred to the two main application classes for CMOS technology – high performance (HP) and low power (LP) – which put different constraints on the static power consumption. At the transistor level, they translate into specifications for the OFF current, *i.e.*, $I_{OFF}$ must be 100 nA/µm in the case of HP devices and 100 pA/µm in the case of LP devices [40][41].

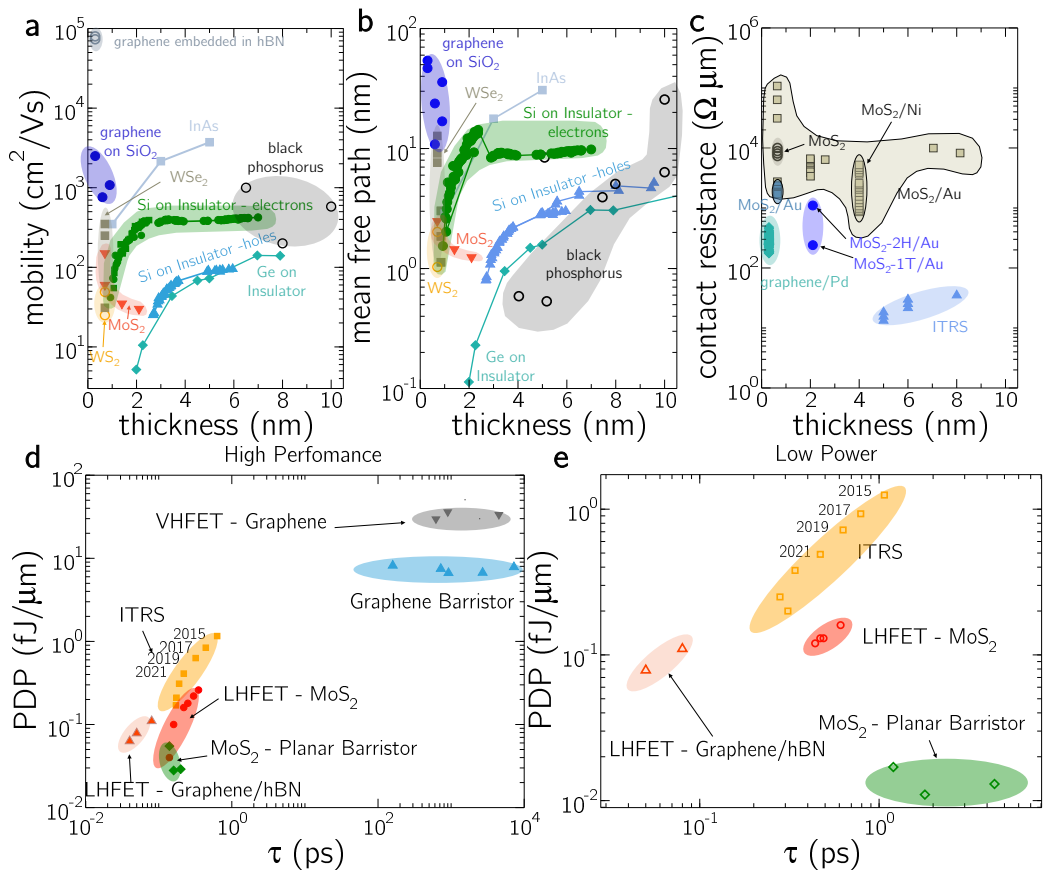

*Figure 4*: a) Experimental mobility vs thickness for different 2D materials and thin body semiconductors. b) Mean free path vs thickness for different 2D materials and thin body semiconductors, extracted from mobility and carrier density measurements. Data sources for a) and b): electrons in SOI [73][74], holes in SOI [74], holes in germanium-on-insulator [75], graphene on SiO$_2$ [76], graphene embedded in h-BN [61], InAs [77], WSe$_2$ [78][79], WS$_2$ [81], black phosphorus [82][83], MoS$_2$ [84][85], c) Experimental Rc of different 2D materials and thin body semiconductors compared with expectations of the ITRS. Data sources: ITRS [40], MoS$_2$ [84], MoS$_2$-2H/Au and MoS$_2$-1T/Au [86], MoS$_2$/Au [87][88] and MoS$_2$/Ni [87], Pd-Graphene [90], d) Scatter plot of delay time and power-delay product for High Performance Logic of different 2D Heterostructure-based FET and comparison with ITRS 2015 e) Scatter plot of delay time and power-delay product for Low Standby Power logic of different 2D Heterostructure-based FET and comparison with ITRS 2015. Data sources for d) and e) considering as a load the gate capacitance of an identical transistor: ITRS [40][41], Graphene-h-BN LHFETs, VH-FET, and Graphene barristor [91], MoS$_2$ LH-FET and MoS$_2$ planar barristor [70].

A lower $I_{OFF}$ for the LP application class typically implies a lower $I_{ON}$ and therefore higher $\tau$ at fixed $V_{DD}$. The route for device optimization is clear: the steeper current modulation the better, since increasing $I_{ON}$ with low $V_{DD}$ and with low C - at fixed $I_{OFF}$ – has a positive impact on both speed and energy efficiency. This requires the barrier potential to be very sensitive to the gate voltage (in the device engineers jargon, the device must have "good electrostatics")[12][41] and the stray capacitance to be minimized [40][41]. Indeed, these two aspects are at the most important differentiators, in terms of application prospects.

Recently, multiscale simulation approaches [91][93][94], ranging from ab-initio simulations of materials and interfaces to complete devices, enabled the quantitative evaluation of the FoM and therefore the optimization of each transistor structure. Here, we choose to follow a *via negativa* approach, and to consider the best-case scenario of defect-less transistors, ideal geometries, and ballistic transport. Transistor structures that even in extremely optimistic conditions would not be competitive with the ITRS requirements should be abandoned as candidate technologies. On the contrary, those that instead are competitive will require some more critical testing and development.

Here, we compare optimized transistor structures based on 2D heterostructures in terms of $\tau$ and PDP for the HP (Figure 4d) and for the LP (Figure 4e) application classes. The ideal regions are the lower left corners. It is apparent that only the lateral barristor and the LH-FETs have competitive FoM that could meet the CMOS technology requirements (ITRS 2.0). All of the vertical devices have worse FoM by three orders of magnitude. The reason is that in the vertical devices there is a semi-metal layer (typically, the graphene layer) that is very close to the gate electrode and that, on the one hand, screens the electric field induced by the gate voltage, worsening device electrostatics and, on the other hand, is responsible for an increased stray gate capacitance. Both effects tend to degrade the FoM values.

## Opportunities and conclusion

Transistors based on heterostructures of two-dimensional materials offer new opportunities to sustain Moore's law, but also pose many challenges. The opportunities reside in the possibility to have atomically thin layers that are only weakly coupled to adjacent ones, providing an easier path to three-dimensional stacking and enabling further scaling of planar transistors with good electrostatics, if stray capacitances are properly minimized. Both conditions are essential to sustain the exponential increase of the number of transistors per chip.

However, 2D semiconductor materials with a semiconducting gap larger than 0.5 eV, required for thermionic FET operation, have poor contact resistance and at best acceptable mobility. Lateral heterostructures can provide a solution by enabling the optimization of materials for different transistor building blocks (regions of semiconducting material for the channel, and quasi-metallic materials for the source/drain regions and contacts). However, growth or formation of lateral heterostructures is still at the infancy stage.

The road for a digital IC technology based on 2D materials therefore cannot be discarded but is extremely narrow. To maximize the probability of success, it is important to focus the efforts on the options that can provide competitive performance at least in the most optimistic case. In our view, such options are represented by planar transistors, using vertical 3D stacking as a further way to improve transistor density and lateral heterostructures to optimize device performance.

## Acknowledgements

We acknowledge financial support from the European Union's Horizon 2020 research and innovation program under grant agreement No. 696656—GrapheneCore1, and a Newton International Fellowship.

## Additional information

## Competing financial interests
The authors declare no competing financial interests.